\documentclass[5p]{elsarticle}
\usepackage{amsmath}
\usepackage{amsfonts}
\usepackage{amssymb}
\usepackage{graphicx}
\usepackage{epsfig}
\usepackage{wrapfig}

\begin{document}
	
	\title{Critical Casimir effects in 2D Ising model with curved defect lines}
	\author[msu]{S.D.~Mostovoy}
	\ead{sd.mostovoy@physics.msu.ru}
	\author[msu,itep,mephi]{O.V.~Pavlovsky\corref{cor1}}
	\ead{ovp@goa.bog.msu.ru}
	\cortext[cor1]{Corresponding author}
	\address[msu]{Faculty of Physics, Moscow State University, Moscow, Russia, 119991}
	\address[itep]{Institute for Theoretical and Experimental Physics, Moscow, Russia, 117218}
	\address[mephi]{National Research Nuclear University MEPhI, Moscow, Russia, 115409}
	
		
		
		
	
	\begin{abstract}
	This work is aimed at studying the influence of critical Casimir effects on energetic properties of curved defect lines in the frame of 2D Ising model. Two types of defect curves were investigated. We start with a simple task of globule formation from four-defect line. It was proved that an exothermic reaction of collapse occurs and the dependence of energy release on temperature was observed. Critical Casimir energy of extensive line of constant curvature was also examined. It was shown that its critical Casimir energy is proportional to curvature that leads to the tendency to radius decreasing under Casimir forces. The results obtained can be applied to proteins folding problem in polarized liquid.
	\end{abstract}
	
	
	\maketitle
	
	\section{Introduction}
	
	The main goal of our work is an investigation of the critical Casimir effect in a number of defects' structures in 2D Ising model. The critical Casimir effect is a back-reaction of the thermal fluctuations of the medium on the factors that have violated the system homogeneity at critical temperatures (phase transition). The effect is an analogy of well-known quantum Casimir effect with quantum fluctuations replaced with thermal ones.
	
	Ising model is a statistical spin model for studying of the properties of ferromagnetic  materials. The main feature of this model is phase transition phenomenon which is connected with spontaneous broken symmetry effect and formation of magnetic domains. The Ising model finds an application not only in material science, but also in other fields of science, for example, biology.
	
	We study vacancies in this paper. These are the simplest defects and are easy for modeling. 
	
	The modern development of the nanophysics gives us the possibility to work with the individual defects. So, they can be used in various nanotechnological applications. Contemporary technology allows to create complicated defect structures \cite{nanoribbons}.
	
	Critical Casimir effect was examined in \cite{fisher,krech}. The effect is of a similar nature as one in quantum field theory, but the latter is caused by vacuum fluctuations. In critical Casimir effect thermal fluctuations of a statistical system act as quantum vacuum ones. The analogy between origin of thermal fluctuation forces and quantum Casimir forces is discussed in \cite{gamb}. The forces are universal and do not depend on microscopic details, so they can be defined by means of scaling functions. This phenomena becomes the most apparent at critical point, where scale invariance takes place. This feature was revealed by Leo Kadanoff in \cite{kadanoff}. That is where the term 'critical' comes from.
	
	The Ising model with bonds changed was studied in \cite{selke1}. Using Monte Carlo, they inspected the correlation functions in two directions as well as the magnetization profile along the lattice. The temperature dependence of the effective correlation length is presented as a function of temperature. Paper \cite{selke2} adds objects, called walls, a set of spins not interacting with neighbors. Results obtained by Monte Carlo and the exact solution are compared. In these articles the Casimir interaction between two lines was considered in particular.
	
	A detailed research on defects in the 3D Ising model is held in \cite{hasen} with usage of a particular cluster algorithm. A system of a spherical object with fixed spins' value and a plate under a range of temperatures was examined. The thermodynamic Casimir force was computed numerically. The author solves the problem of discretization of a sphere, which we also do in this paper in case of an arc of a curve. The efficiency of numerical computations is discussed in detail. They checked how results change with lattice volume increase and concluded that ``in a neighbourhood of the critical point the effect of the finite lattice size on energy is clearly visible'' \cite{hasen}. We are aware of this problem, so we considered a set of volumes (from 20 to 160) in order to find out infinite volume limit.
	
	Casimir effect emerges in various fields of physics. For example, critical Casimir forces in a biophysics application are calculated analytically in \cite{nowakowski}. Membranes of living cells exhibit properties similar to those of the critical Ising model. In this paper protein inclusions in cellular membrane are modeled by modifying a set of coupling constants in infinite Ising lattice, and critical Casimir forces are computed analytically by transfer matrix diagonalization. It should be noted that in our work we modify \textit{all} the coupling constants around a selected spin. The Free energy of two interacting defects is evaluated in \cite{nowakowski} in thermodynamic limit. One of the results, important for our research, is the attraction of \textit{identical} defects, which is mentioned in the paper considered and known from an exact solution for the planar Ising model on the square lattice \cite{hecht}. The existence of the attraction is the foundation stone in our discourse.
	
	Another approach to the study of the properties of critical Casimir effect in biophysical problems is developed in work \cite{cells}. The authors of the article proposed that this ensures the interaction of biological structures by Casimir forces at large distances. They calculate the interaction forces potential of membrane inclusions. Calculations performed within the framework of conformal field theory are verified by numerical calculations by the Monte Carlo method. It is shown that fluctuations in the composition of plasma membranes of living cells appear in the occurrence of long-range forces acting on proteins included in membranes. It is assumed that the critical Casimir forces serve for the initial stages of signal transmission. It is shown that the forces can be attractive or repulsive depending on the type of inclusions involved in the interaction. Three approaches were used to evaluate the potentials. Firstly, from thermodynamics, by using a partition function for the combined system of the Ising model plus proteins (point-like defects). Next, numerically, using Monte-Carlo on the lattice Ising model for like and unlike	disk-shaped inclusions. Finally, conformal field theory helped to make an analytical	prediction for the form of the potential. This CFT approach was used in \cite{EB} for the complete \textit{analytic} treatment of defects led to proof of the attraction of point-like and disk-shaped defects and to understanding of the role the boundary conditions play.
	
	Besides, the lattice models are applied in studying the properties of various media with characteristic features, for example, water. Water has a significant impact on the physical and biological processes that occur in the medium. Various models of water are widely used to study this effect. Lattice gas models are particularly convenient to use in theoretical and numerical studies. This model was first proposed in \cite{bell1, bell2}. Within the framework of these models, water is considered as a lattice statistical system where the nodes are water molecules. This consideration does not take into account the dynamic processes associated with the movement of water molecules, but it simplifies the calculations considerably. The objects under the investigation, placed in the medium, can be considered within the framework of the water lattice model as a set of defects (vacancies) which displace water from its volume. In \cite{titov} Monte Carlo numerical studies of 3D water properties are represented in the Lattice gas model that was suggested in two papers of G. M. Bell and co-workers.
	
	A water molecule has two positive and two negative polarizations in 3D. These polarizations make up a tetrahedron in space. In a simplified 2D investigation, one disregards the 3D structure of the water molecule polarizations. The sum of positive and negative polarizations is summed up in pairs into one dipole moment for each water molecule, and the complex interaction of such models is approximated by the simple interaction of such dipoles.
	
	The use of Ising model with defects in biophysical applications could be relevant in cases when medium polarization effects must be taken into account. $\rm H_2O$ molecules are polarized and thus could be modeled by spin lattice in such a way as ferromagnetic materials due to the same role of dipole molecule moment as magnetic moment of substance elements. For example, the Ising model is applied to description of water properties in nanopores \cite{nanopores}. Historically, it was Lee and Yang who used the lattice gas model as an aid to comprehension of the liquid-gas phase transition and its critical point \cite{lee}. A similarity between real magnetic transitions and real liquid-gas transitions is demonstrated.
	
	It is known that temperature of the phase transition in 2D Ising model depends on the defects' presence. A thorough research of this phenomenon is enclosed in \cite{os}. They have showed, that the critical temperature variation is proportional to concentration of defects defined as ratio of the number of defect to the total nodes. In \cite{ha} the results presented in \cite{os} are refined by the cumulant expansion and generalized to 3D Ising. A polynomial dependence of the critical temperature on the concentration of defects is obtained. In our paper we will consider the structures of defects, whose number grows slower than the volume does, and therefore, based on the results of the paper \cite{os}, the variation of the critical temperature value can be neglected for large volumes.

	Vacancies grouped in specific formations -- lines and globules -- are studied in our work. The Casimir force appears to change initial form of defect structures. In our investigation we use the numerical Monte Carlo technique.
	
	\section{Definitions}
	\label{sec:def}
	
	Let us introduce a 2D lattice of square unit cells with $L$ nodes along each side (so, the lattice contains $L^2$ nodes in total). In the paper $L$ varies from 20 to 150. We study vacancies that are absent atoms in certain nodes. The defects are put in corresponding places and the total energy is calculated by means of Monte-Carlo modeling. In Ising model the configuration energy is
	\begin{equation}
	\label{eq:energy}
	E(Conf) = -J\sum_{x,\mu}\sigma_{x}\sigma_{x+\mu},
	\end{equation}
	where $J$ is the coupling constant and $\sigma_{x}$ is a spin of node $x$. $x$ enumerates all the nodes. $\mu$ is a shift to one column to the right or one row below, so that all the neighboring pairs of nodes are involved once in the sum. In this calculation $J=1$ is used. Although $J$ in (\ref{eq:energy}) is written without indices, it does depend on $x$ and $\mu$. Actually, to put a vacancy in position $x_0$, all the four links around $x$ are crossed out by setting appropriate $J$ to zero. While simulation, it is sufficient to exclude four terms from the summation (\ref{eq:energy}). A single defect is just a vacancy, while an extensive one is a group of adjacent vacancies with all corresponding $J=0$.
	
	Periodic boundary conditions are applied. (Therefore, if $x$ in (\ref{eq:energy}) is in the rightmost column or the bottommost row, $x+\mu$ will be in the leftmost column or the topmost row, respectively). Our aim is an examination of critical Casimir effects which are energy effects over pure lattice (without defects) characteristics. So, we determine the internal energy of defects as the contribution of defect's presence to the full lattice energy:
	
	\begin{equation}
	\label{eq:inten}
	E_{\mbox{def}} = E - E_0,
	\end{equation}
	where $E$ is full energy of lattice with defects (\ref{eq:energy}) and $E_0$ is the same but for regular lattice. Thus, we have selected a reference level of energy.
	
	It is worth pointing out that individual defects were examined in \cite{china}. It was proved that defect's internal energy has a peak at critical temperature of Ising model $1/T_c \approx 0.44$, and the maximum rises with lattice volume. This is why it is essential to compare configurations with equal defects' quantity in every simulation in order to avoid the emergence of divergences.

	We determine a defect line by putting defects next to each other. No interaction between the spins through the defect line occurs. (it was just mentioned when we set $J$ to zero.) So we say that a defect breaks down contact between nodes completely. There are no bindings both within the line and between a spin within the line and the surrounding spins non included in the line.
	
	During the numerical calculation the defects are fixed in their places, but by examining two or more slightly different configurations of defects we can determine dynamic tendencies in the system. This means that due to Minimum total potential energy principle the system considered had changed its state in a specific direction if we studied its evolution. However we deal with a set of fixed defects configurations which can not turn to each other. So, only \textit{tendencies} in change of configurations can be noticed.
	
	As usual for classical systems, Boltzmann distribution is used, though we assume Boltzmann constant $k_B=1$, so the probability of a configuration is
	$$
	P(Conf) = \frac{1}{Z}e^{-E(Conf)/T},
	$$
	where $Z$ - normalization factor.
	
	\section{Collapse of four-structure object}
	\label{sec:globule}
	
\begin{figure}[h]
		\centering
		\begin{tabular}{ccccc}
			\includegraphics[width=0.16\linewidth]{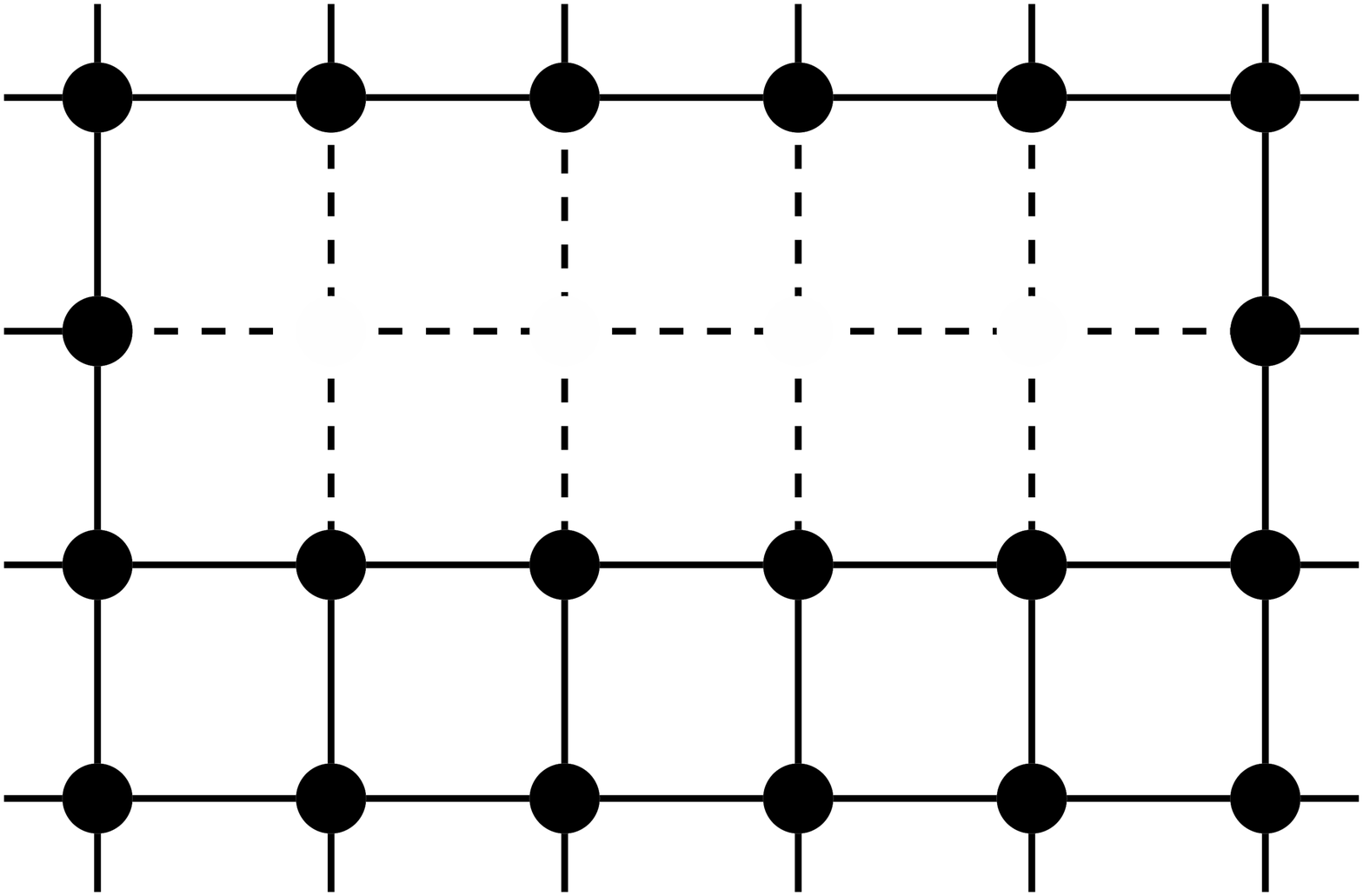}&\includegraphics[width=0.16\linewidth]{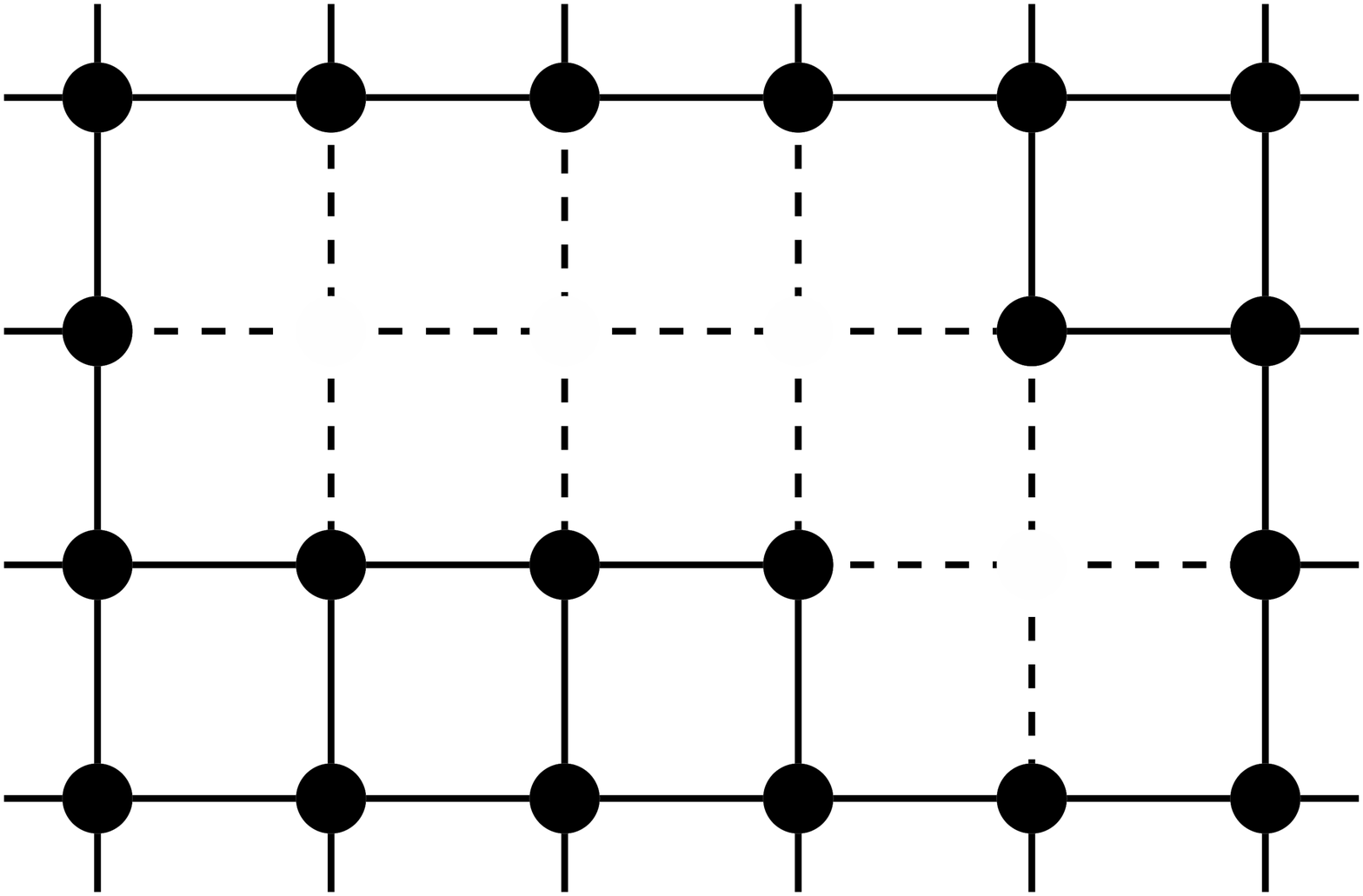}&\includegraphics[width=0.16\linewidth]{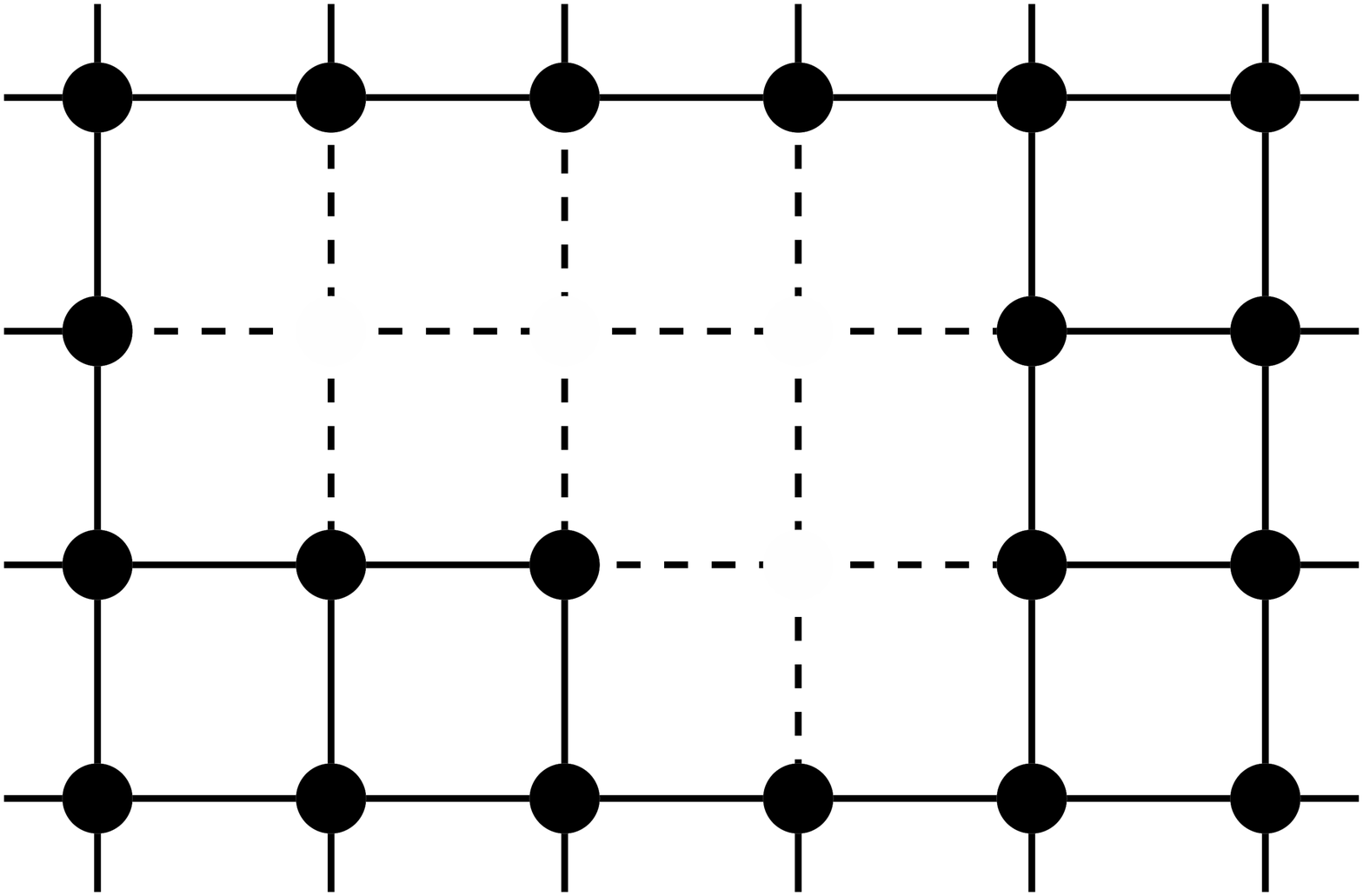}&\includegraphics[width=0.16\linewidth]{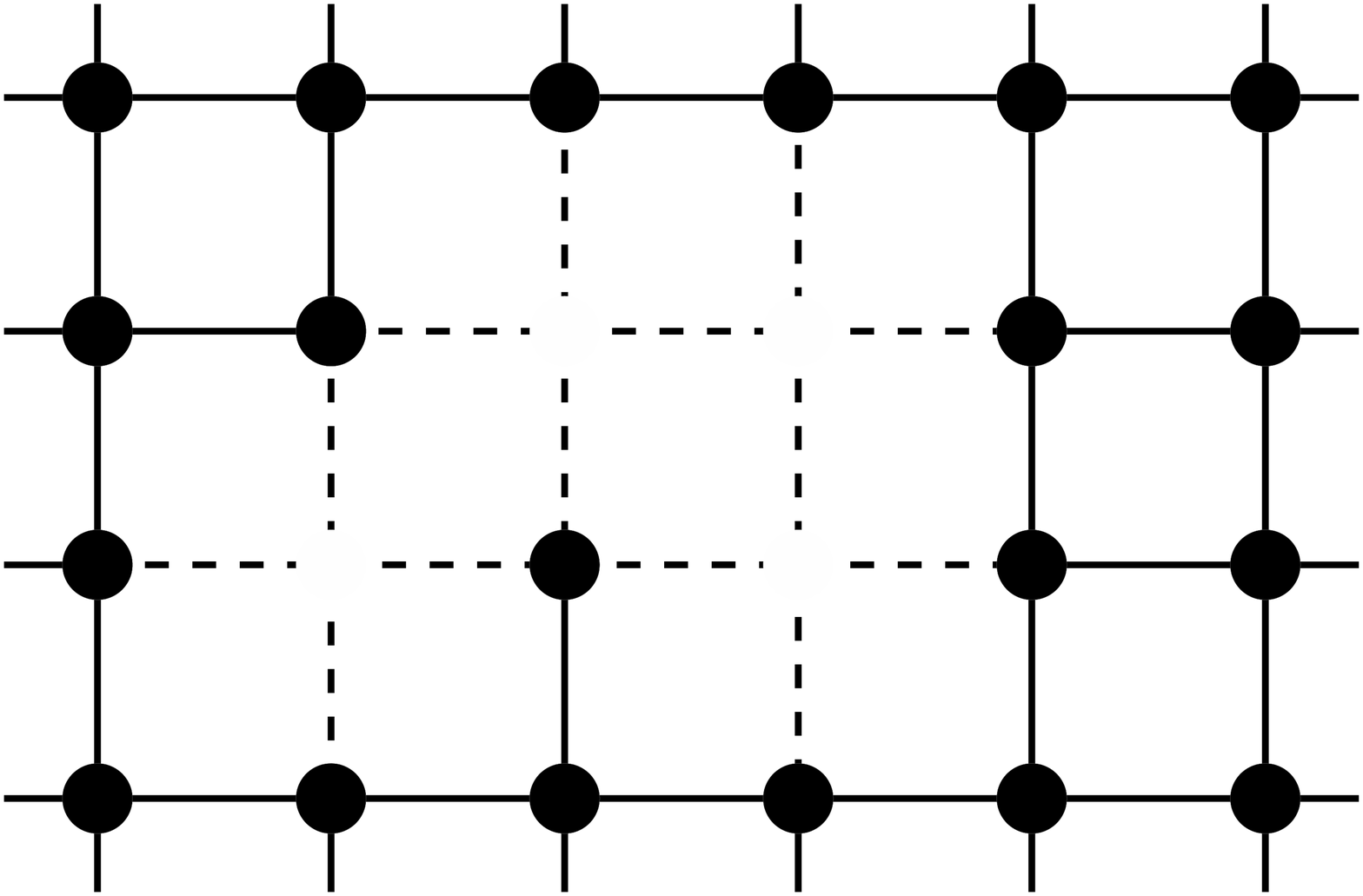}&\includegraphics[width=0.16\linewidth]{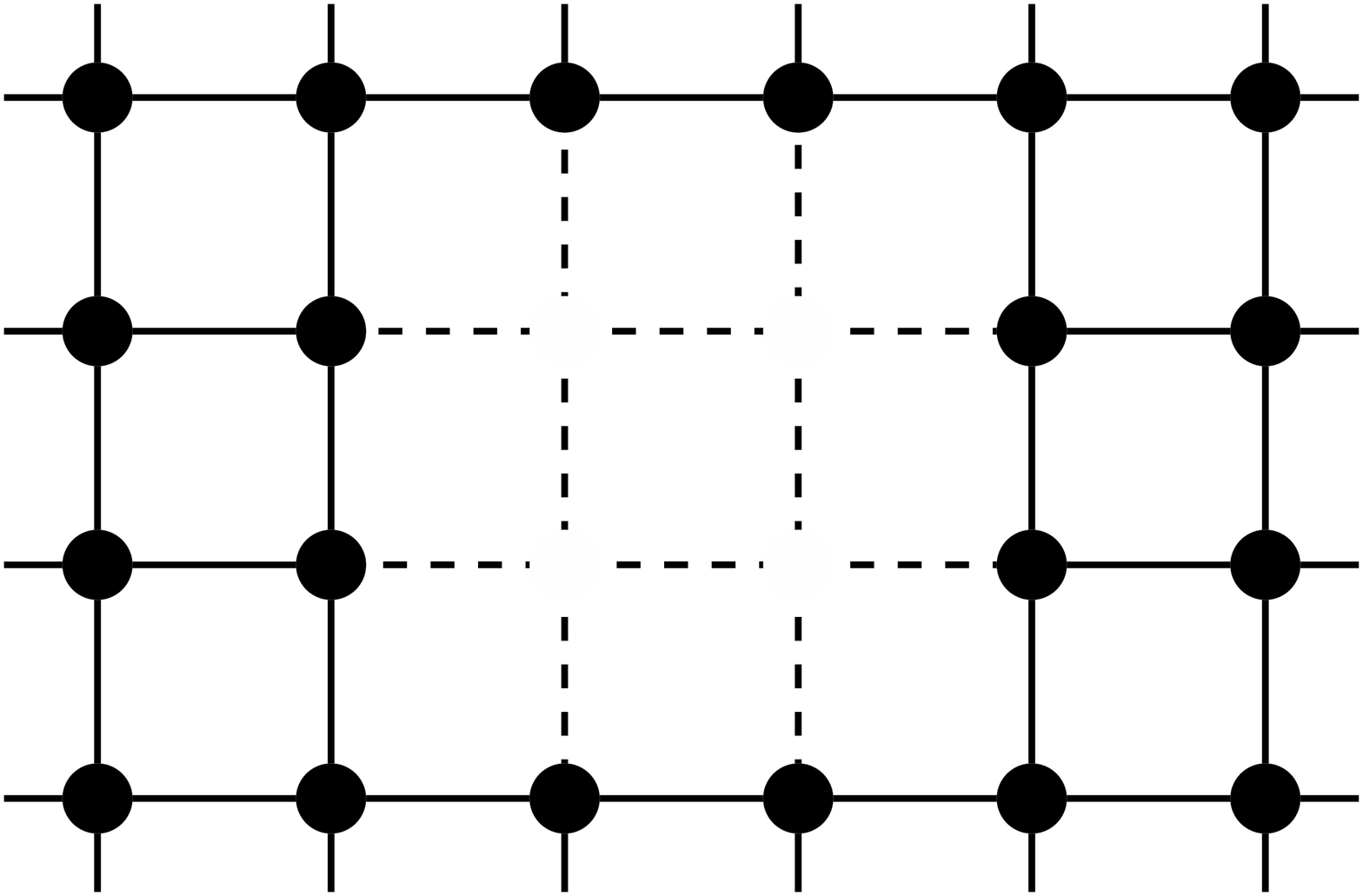}\\
			a&b&c&d&e\\
		\end{tabular}
		\caption{A series of configurations to form a globule from line. Method of links counting (see main text) helps to estimate relation between energies of configuration pairs.}
		\label{fig:globula_confs}
	\end{figure}

	Let us turn to the distortion of 4-nodes long defect line into a globule --- a square-shaped group of four defects. This makes it possible to draw an analogy between the biophysical process of formation of the tertiary structure of the protein and the formation of dense defect structures in the model of theoretical physics. For the Ising model the energy properties of defects are also determined by their \textit{environment}, namely by the distribution of the neighboring nodes. This simplest object has been chosen because a longer line can be twisted in a wider variety of patterns. So, it is not clear how to choose an appropriate sequence of configurations. In our case, the decision is almost evident: see \figurename~\ref{fig:globula_confs}.
	
	We are interested in the energy difference between successive configurations of the arrangement of defects in our sequence (see \figurename~\ref {fig:globula_confs}). The defects are fixed in their positions, which allows us to  calculate the total energy of the system by Monte Carlo simulation.
	
	The calculations below were fulfilled with CUDA technology by using graphics cards. We employed multithreading approach to thermalization via thermal bath method using checkerboard algorithm \cite{chboard}. The structure of the Hamiltonian (\ref{eq:energy}) allows us to distinguish two sublattices nested inside each other in a square lattice. These sublattices do not interact at each step of the thermalization. As a result, $L^2/2$ CUDA calculation threads, changing each other, handle both the sublattices. The algorithm leads to a rapid change of the system states, which reduces the total number of required thermalization steps. The value of the total energy is calculated after a certain number of steps and then an average is found. In addition, statistics is improved through the usage of several independent thermalization sequences.
	
	\begin{figure}[h]
		\centering
		\includegraphics[width=0.8\linewidth]{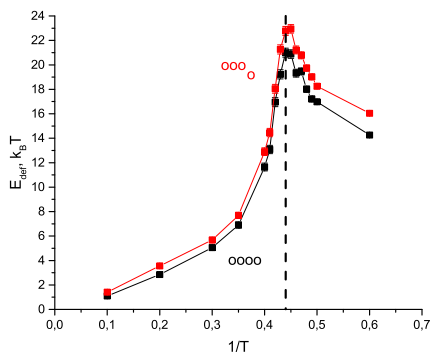}
		\caption{Internal energy of 4 defect line in a square $20\times20$ lattice as a function of inverse temperature. The peak is reached near critical temperature (denoted by dashed line). Here and below a system of units with $k_B=1$ is used.}
		\label{fig:globula_lineen}	
	\end{figure}
	
	In \figurename~\ref{fig:globula_lineen} internal energy $E_{\mbox{def}}$ (\ref{eq:inten}) of a straight line (\figurename~\ref{fig:globula_confs}a) vs. inverse temperature is shown. The curve's peak can be understood as a consequence of scale invariance at phase transition\footnote{See accurate derivation of $T_{cr}$ in \cite{critCasTemp}. For our calculation only two digits are used due to applying a grid in 0.01 steps.} ($1/T_{cr} \approx 0.44$). In fact, peak's value is limited only by finite size of the lattice in numerical simulation which can be shown by calculation. This is so, because at critical point distances at which spins ``feels'' each other are very long (they can exceed the lattice size). Therefore, the defect affect the whole lattice. The effect considered is of interest in vicinity of the critical temperature, so inverse temperatures $0.3$, $0.4$ and $0.44$ have been chosen.
	
	The main question is how the system of 4 defects acts. Is there a stability state? Or does a process of collapsing take place? Would this process go to the end? We suppose that the answer can be obtained while estimating the threshold energy. In fact, an analysis of the reaction path (a sequence of states of the system through which the transformation takes place from the initial state to the final state) is considered in the course of physical chemistry. It is demonstrated that the only obstacle to the spontaneous flow of the reaction is the energy barrier. Number of such barriers and their height determine the process probability. The matter is:
	
	\begin{enumerate}
		\item If the final energy is less then the initial one, the reaction is exothermic and begins spontaneously due to the lowest energy principle.
		\item If some local maximums on reaction path present, the probability of transition decreases. Additional energy is essential for the reaction flow.
		\item Different circumstances may change the energy profile varying the reaction possibility. For example, a catalyst reduces the energy barrier of the reaction. In some cases, the atoms in the excited state react more readily.
	\end{enumerate}
	
	Turning back to the Ising model, it is worth saying that we \textit{may} predict the behaviour. The first reason is the number of links between defects. The total energy (\ref{eq:energy}) increases if a link is stricken out. However, energy reduction is of our interest. Let us count the number of links in two situations: the straight line and the final globule. This count recedes from 13 to 12. A reaction \textit{may be} spontaneous. However in intermediate states (\figurename~\ref{fig:globula_confs}b and 1d) this count rises up to 14. We observe two barriers: while transition from state 1 to 2 and from 3 to 4.
	
	\begin{figure}[h]
		\centering
		\includegraphics[width=0.2\linewidth]{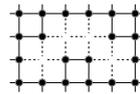}
		\caption{Another possible defect configuration: both sides of the linear defect start to fold. The link counting technic shows that this state is unlikely.}
		\label{fig:globula_lat6}	
	\end{figure}
	
	Now, it becomes clear, why we do not consider a configuration depicted in \figurename~\ref{fig:globula_lat6}: actually, 15 links are stricken out. The internal energy increases accordingly. Provided the lattice is in configuration \textit{b} (refer \figurename~\ref{fig:globula_confs}), the system is likely to go into state \textit{c} rather than the one in \figurename~\ref{fig:globula_lat6}.
	
\begin{figure}[h]
		\centering
		\includegraphics[width=0.9\linewidth]{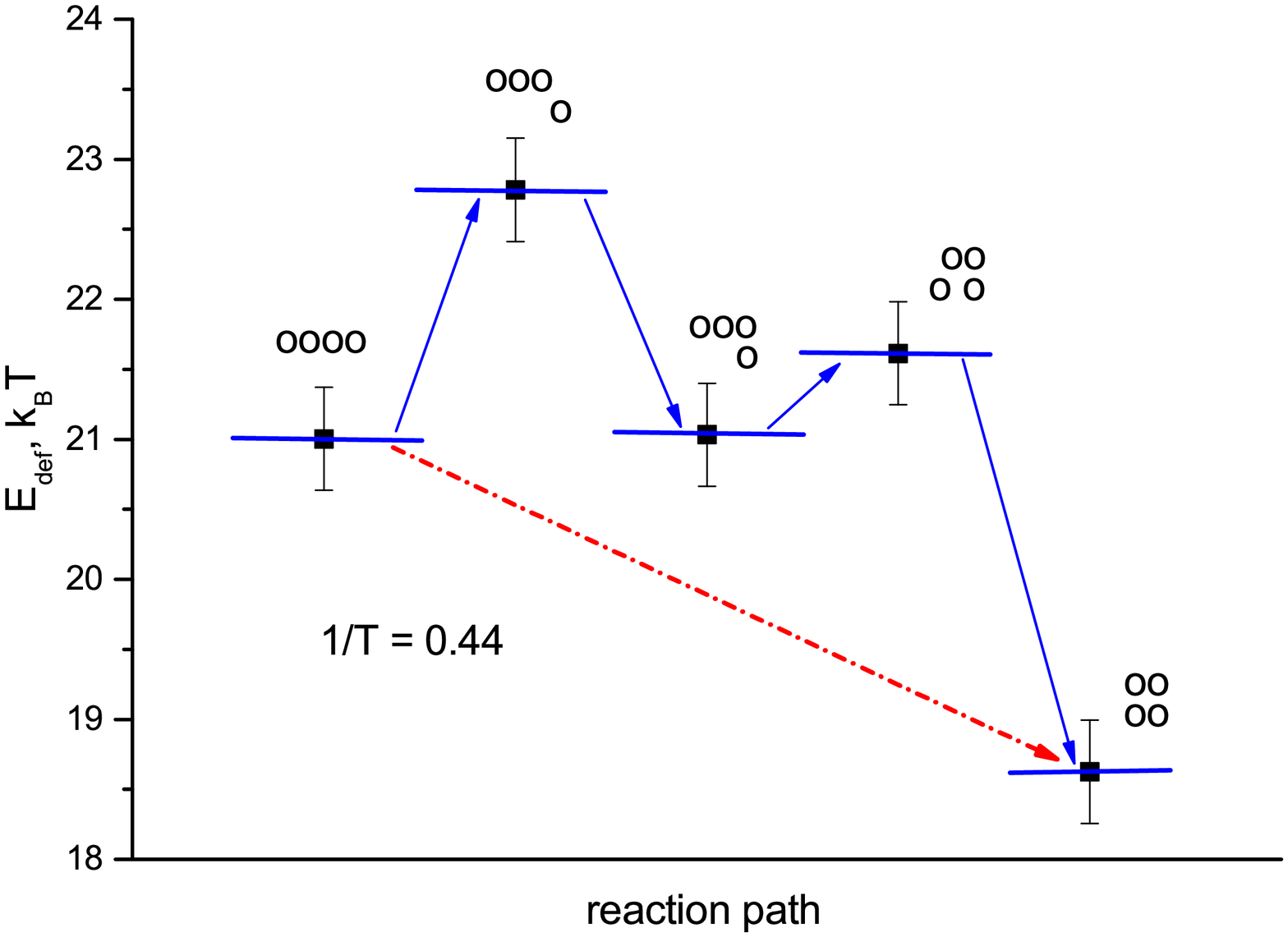}
		\caption{Reaction path of 4 defect line collapse in a $20\times20$ lattice. The blue solid arrows show the energy consumption at every step, the red dashed one is the energy release of the whole process. Apparently, the collapse is an exothermic transformation.}
		\label{fig:globula_energy44}
	\end{figure}

	Let us check our prediction with direct Monte Carlo simulation. The graphs of defects' internal energy (\ref{eq:inten}) at $1/T=0.44$ is shown in \figurename~\ref{fig:globula_energy44}. We see the confirmation of our statement:
	
	\begin{quote}
		counting of links determines the most preferable defect configuration if number of defects is fixed.
	\end{quote}
	This statement will be used in this paper to estimate the system behaviour. \figurename~\ref{fig:globula_energy44} shows that collapsing is beneficial for the system, however the process is not spontaneous: an activation energy required to trigger the transition. The emergence of the barriers is connected with the necessity to break and establish links between nodes.
	
\begin{figure}[h!]
		\centering
		\includegraphics[width=0.5\linewidth]{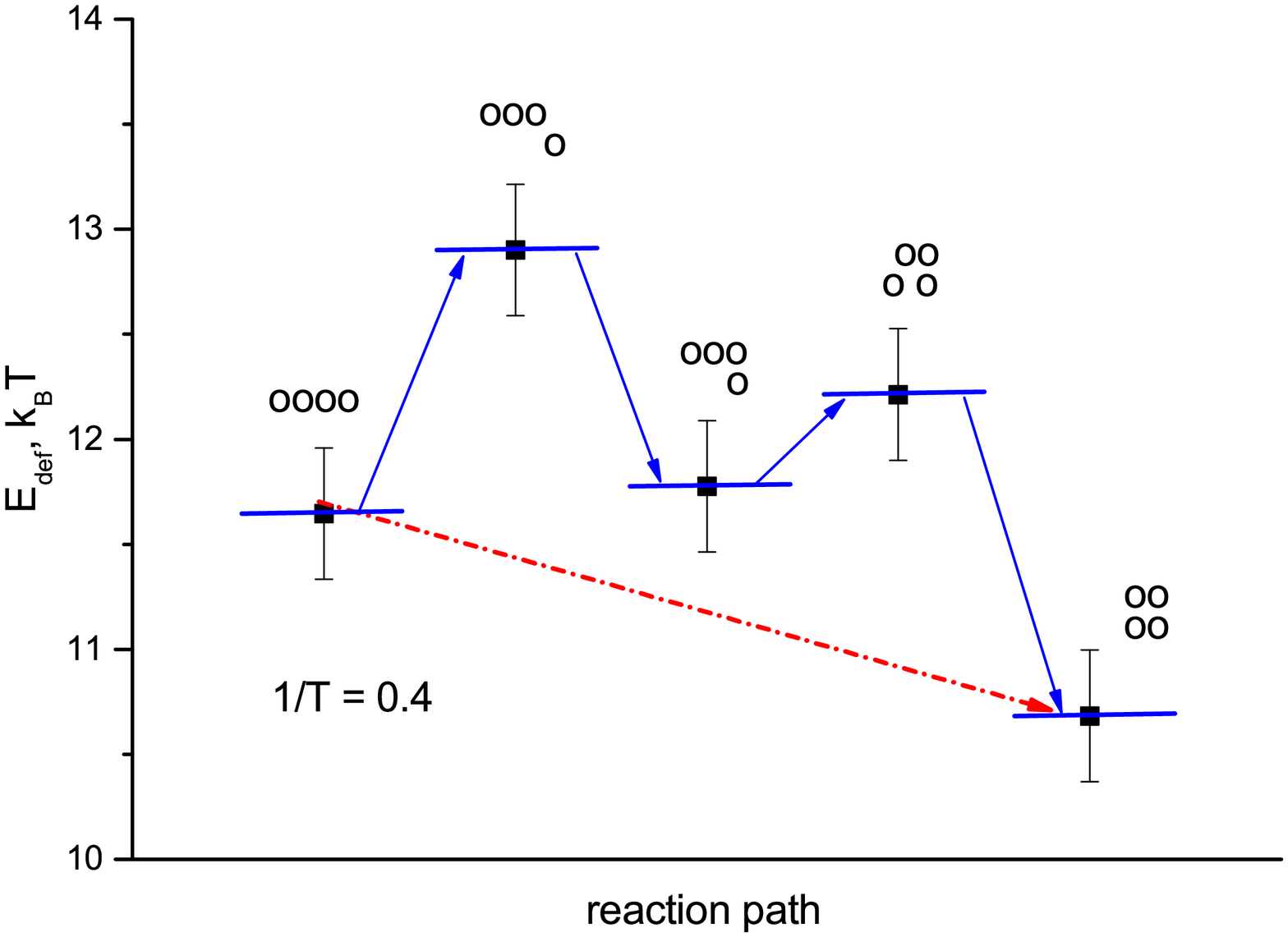}\includegraphics[width=0.5\linewidth]{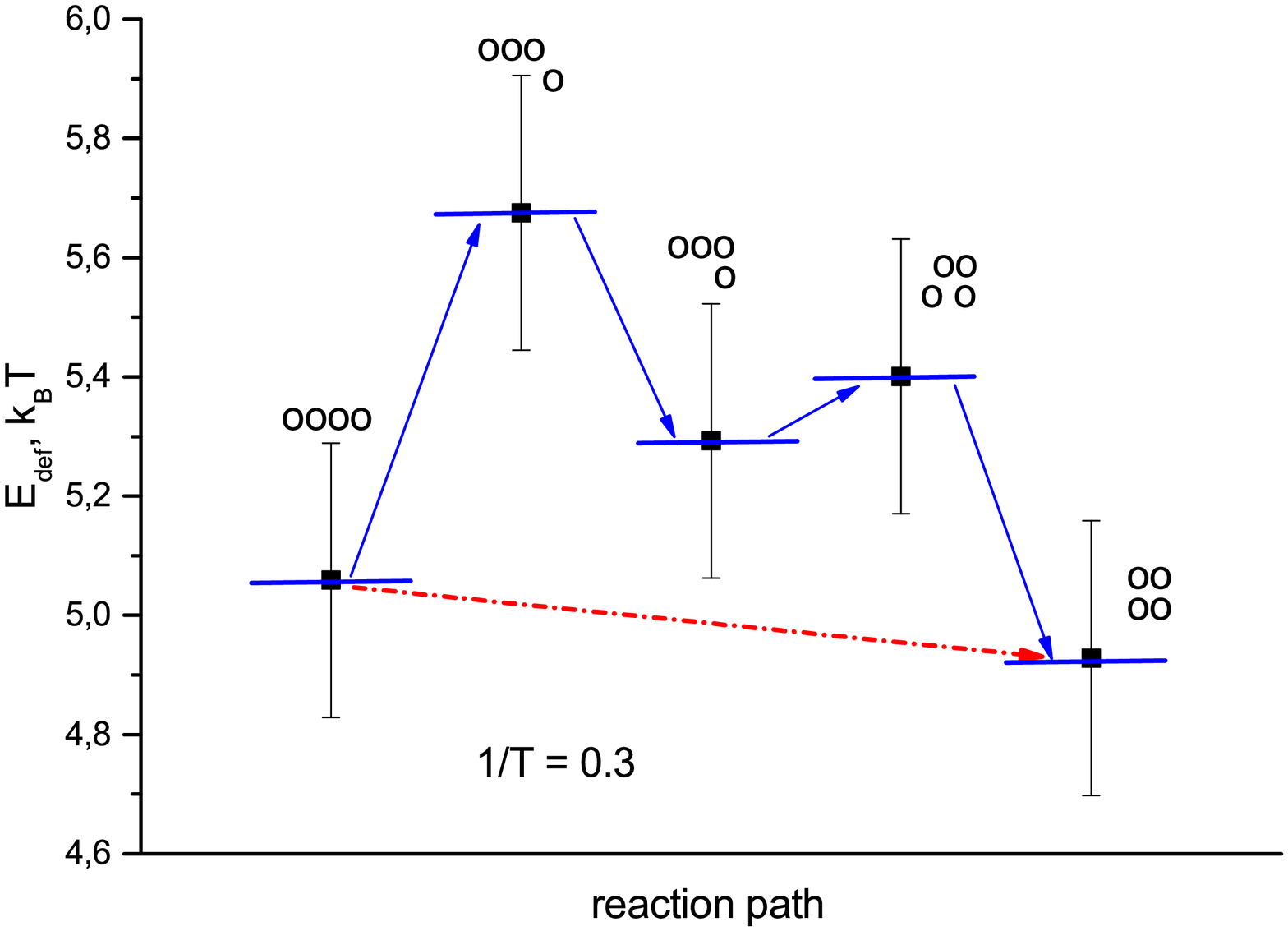}\\		
		\caption{Energy release of collapse for temperatures above phase transition. The benefit is ambiguous, so the process becomes unlikely.}
		\label{fig:globula_energys}
	\end{figure}

	In our work we also have shown that energy magnitudes of conversions are the greatest at critical temperature. As an evidence turn to \figurename~\ref{fig:globula_energys}. Notice that energy release is about $2.4$ at $1/T=0.44$ while it is $1.0$ at $1/T=0.4$ and $0.13$ at $1/T=0.3$. Therefore, in the high temperature phase the collapsing faces a significant barrier, but benefit is rather poor.
	
\begin{figure}[h!]
		\centering
		\includegraphics[width=0.5\linewidth]{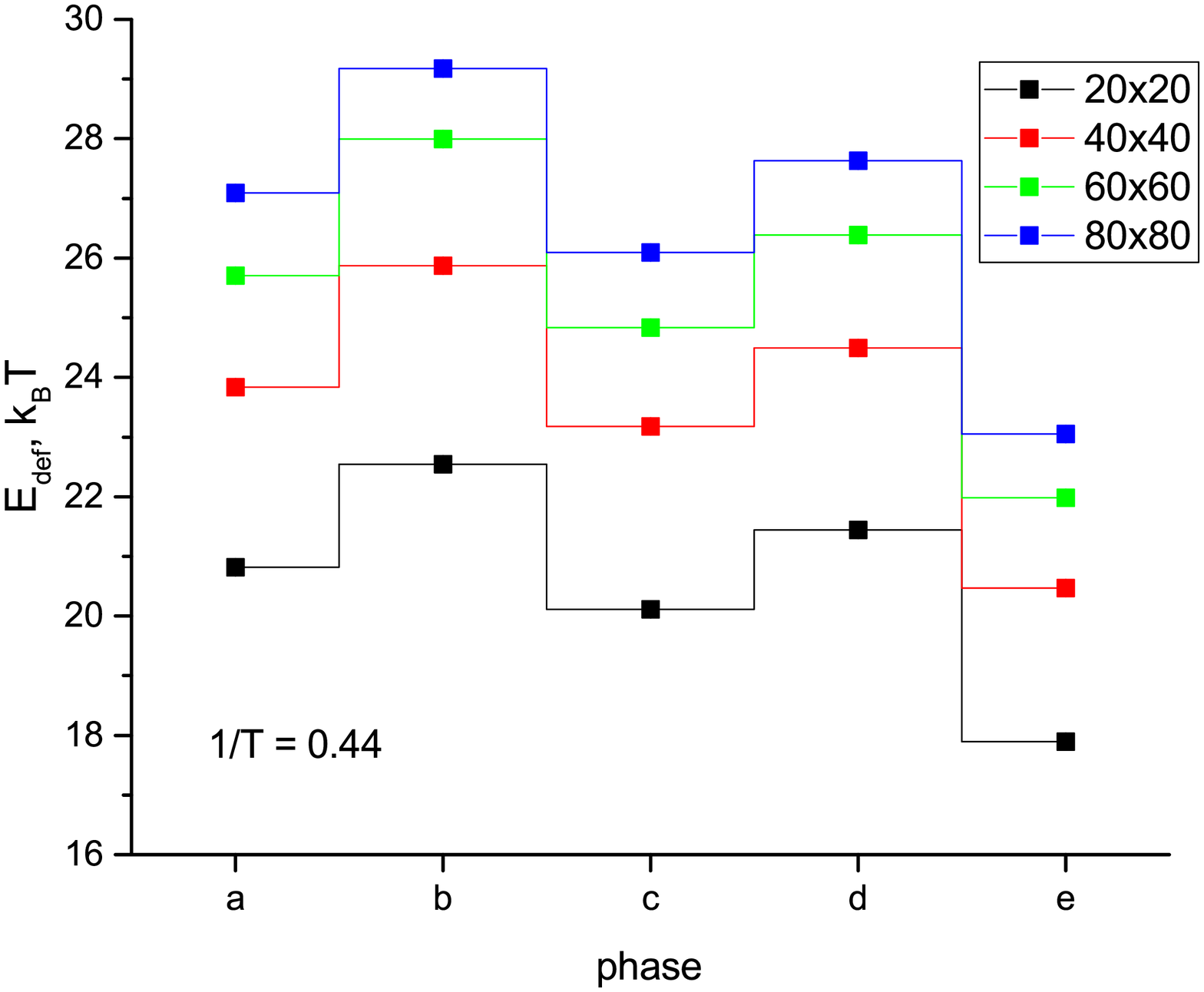}\includegraphics[width=0.5\linewidth]{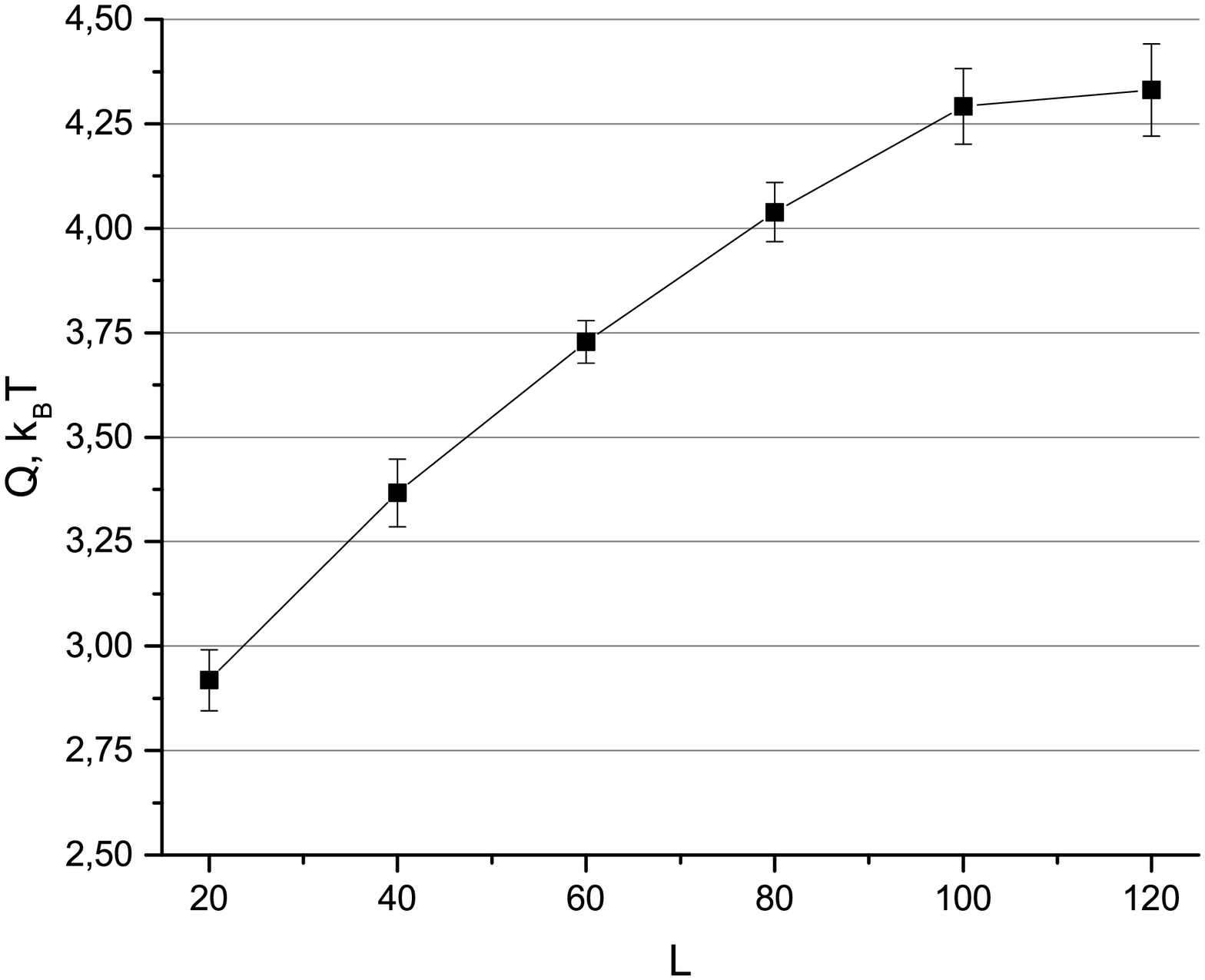}\\		
		\parbox{0.48\linewidth}{\caption{Energy of the phases (see. \figurename~\ref{fig:globula_confs}) for the line collapse in different lattices. All the curves are similar, so the effect exhibits the features mentioned in lattice of any volume.}
			\label{fig:globula_energy44vol}}\hfill
		\parbox{0.48\linewidth}{\caption{Energy release $Q$ as function of the lattice size. The saturation achieved gave us the possibility to predict the value of $Q$ in an infinite volume limit ($L\to\infty$).}
			\label{fig:globula_qvol}}
	\end{figure}

	Hitherto, we have considered a lattice with size of $20\times20$. As we know, volume effects can change the observables dramatically, so we should examine lattices of a larger size. The study discussed so far reveals the main features of the phenomenon, so it appeared to be productive to work with such a small lattice since few technical problems occur. \figurename~\ref{fig:globula_energy44vol} shows that the nature of the collapse does not change, then we deal with large $L$, where $L$ is the edge of the lattice. So, we claim, that all the results obtained in a $20\times20$ lattice are valid in an infinite volume limit (where $L\to\infty$) also.
	
	A red dashed line designates a reaction energy release $Q$ in \figurename~\ref{fig:globula_energy44}. This quantity is defined as
	\begin{equation}
	Q = E_{\mbox{def}}(a) - E_{\mbox{def}}(e),
	\end{equation}
	where $E_{\mbox{def}}(x)$ is the internal energy (\ref{eq:inten}) of a defect configuration, denoted as $x$ (refer to \figurename~\ref{fig:globula_confs}). $Q$ is positive, so the reaction is exothermic. The volume dependence of $Q$ is depicted in \figurename~\ref{fig:globula_qvol}. It is clear that $Q$ is finite and has a certain limit for $L \to \infty$. That makes physical sense and thus this quantity is an observable whereas the reaction could be accomplished in practice.
	
	\section{Curved defect line}
	
	Let us consider a defect line as large as the lattice size. As shown above, defects tend to attract to each other, which can lead to collapse of an extensive object (see Section \ref{sec:globule}). This time, we do not want to break the line, but only twist it. As a result, a number of possible defect configurations increase tremendously. To characterize the resulting curves, we use a notion of \textit{curvature} $K$: all the 2D curves form arcs of a certain radii with the constant length.
	
	\begin{figure}[h]
		\centering
		\includegraphics[width=0.25\linewidth]{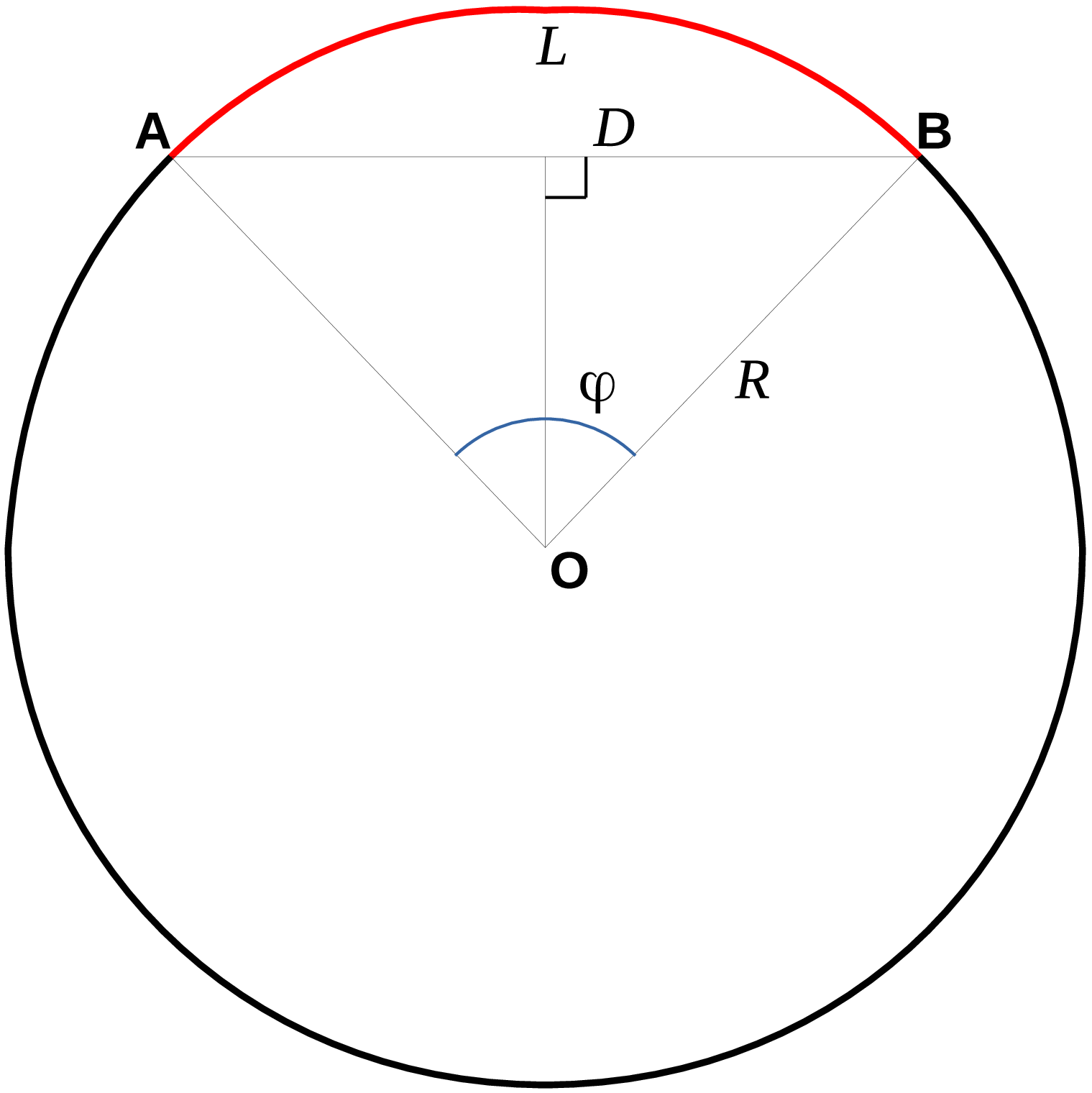}
		\caption{An isosceles triangle $AOB$ helps to find out $AB$ if angle $\varphi$ is given.}
		\label{fig:circlegeom}
	\end{figure}
	
	To begin with, we expect something like a collapse of the curved defect line due to the mutual attraction of the defects. Let us show this analytically. We find the average distance between the points of an arc of radius $R\in\mathbb{R}$. \figurename~\ref{fig:circlegeom} shows that the distance between end points of arc of a circle with central angle $\varphi$ is equal to
	$$
	D = 2R\sin{\frac{\varphi}{2}}.
	$$
	For a fixed length of the circle arc $L$, we chose a set of values for radius $R$. In each case an arc of circle with the angular size $\varphi_m=L/R$ is examined. It is worth to determine the average distance from one end of the arc ($\varphi_0 = 0$) to all its points as a function of $L$ and $R$:
	\begin{equation}
	\label{eq:avecircleintenergy}
	\langle D\rangle = \frac{1}{\varphi_m}\int_0^{\varphi_m} D(\varphi) \,d\varphi = \frac{4R^2}{L}\left(1-\cos\left(\frac{L}{2R}\right)\right).
	\end{equation}
	
\begin{figure}[h]
		\centering
		\includegraphics[width=0.5\linewidth]{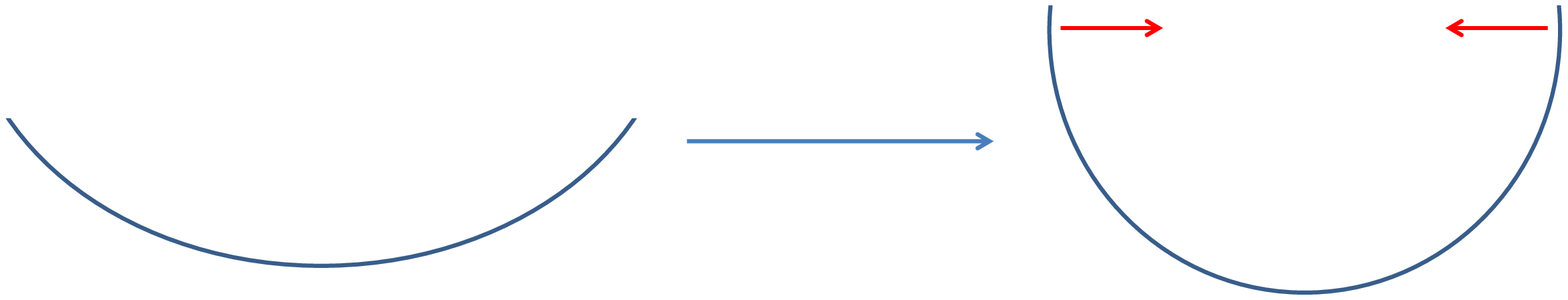}
		\caption{A ``collapse'' of a line into a semicircle. Casimir forces (red arrows) guide the process.}
		\label{fig:curved_arccollapse}
	\end{figure}

	For $L$ fixed $\langle D\rangle$ grows monotonically with $R$. Given that the defects are attracted to each other, we come to the conclusion that the curve tends to contract in a circle (see \figurename~\ref{fig:curved_arccollapse}).
	
	\begin{figure}
		\centering
		\includegraphics[width=0.5\linewidth]{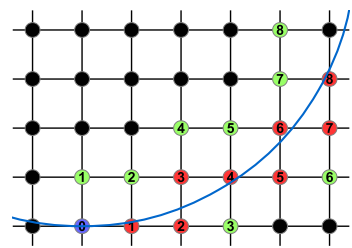}
		\caption{Discretization of a defect curve (blue). Red denotes the approximating lattice nodes.}
		\label{fig:curved_discr}
	\end{figure}
	
	To form defect configurations, we need to convert a smooth curve to a discrete model that will allow us to identify the nodes occupied by defects. To do this, consider the following \textit{natural} approach. Assume that the radius $R\in\mathbb{N}$ of the arc is equal to an integer number of lengths of lattice constants. Fix the node above which the lowest bottom point of the supposed circle will be located (node 0 in \figurename~\ref{fig:curved_discr}). Then two adjacent nodes are considered: to the right and top. The distances from these nodes to the node, where the center of the circle is, are measured and the node that provides the minimum of the deviation is selected:
	
	$$
	\min_j\,\lvert r_j - R\rvert,\qquad j=1,2,
	$$
	where $r_j$ is the distance from the center of the circle to the $j$-th node. \figurename~\ref{fig:curved_discr} demonstrates the successive steps of selecting sampling points, labeled with numbers from 1 to 8. The total amount of defects to be put in the lattice is determined by the line length in an infinite $R$ limit (when the line is straight). So, we just put defects as described above until total number $L$ is reached. This was done in order to fulfill the requirement mentioned in Section \ref{sec:def}: a number of defects should stay constant. As a drawback, the defects' systems generated by the algorithm not always look like a smooth arc of a circle.
	
	Let us check the advantage of turning a line into an arc of a circle numerically. For this purpose, we choose lattices of several sizes (20, 75, 100 and 150) and fill them with the discretized arcs of defects with different radii. With the help of Monte Carlo simulation, the total energy of the system was calculated. All the computations were done at $1/T=0.44$. In paper \cite{os} it is shown that critical temperature depends on defects' concentration. However, as we consider curves of length $L$ in lattices of size $L^2$, the concentration decreases as a function of $L$. In such a situation, we neglect a small difference between $0.44$, the critical temperature of pure lattice and the bulk critical temperature of the system considered. After we subtract the total energy of the configuration corresponding to a certain large radius (an immediate check showed that there is no difference, to do a subtraction with a straight line or with a slightly curved arc). So, the energy $E$ depicted in the figures of this Section is the following:
	
	\begin{equation}
	E = E_{\mbox{tot}}(K) - E_{\mbox{tot}}(K_0),
	\end{equation}
	where $K_0$ is a constant (rather small) curvature and $E_{\mbox{tot}}$ is the total energy (\ref{eq:energy}).
	
	Let's pay attention to the external parameter included in our consideration: the size of the lattice. As shown in Section \ref{sec:globule}, the difference in energy due to link number change gives a meaningful difference only if the number of defects is unchanged. However, by construction, the length of the curve is directly proportional to the size of the lattice. Therefore, later in this section we \textit{normalize} the energy values by the lattice size (estimating the energy per 1 link).
	
\begin{figure}[h]
		\centering
		\includegraphics[width=0.95\linewidth]{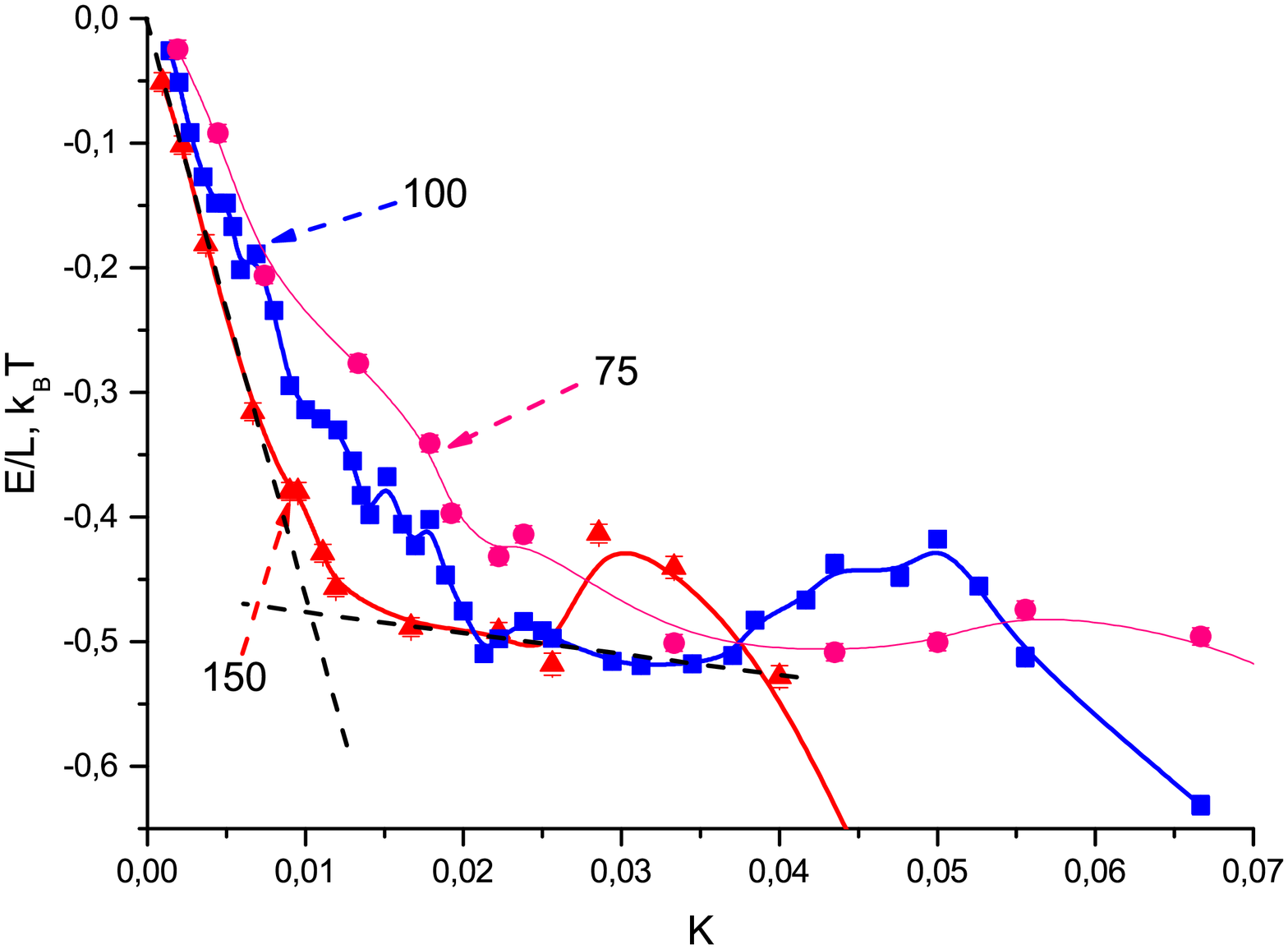}
		\caption{Energy per unit length of the curve as a function of the size of the lattice and curvature. A curve is as long as the lattice size. Attraction prevails, not counting the local repulsion area. The reason for the effect is explained in the text. The solid line connecting data points is a B-Spline.}
		\label{fig:curved_energy1}
		\includegraphics[width=0.95\linewidth]{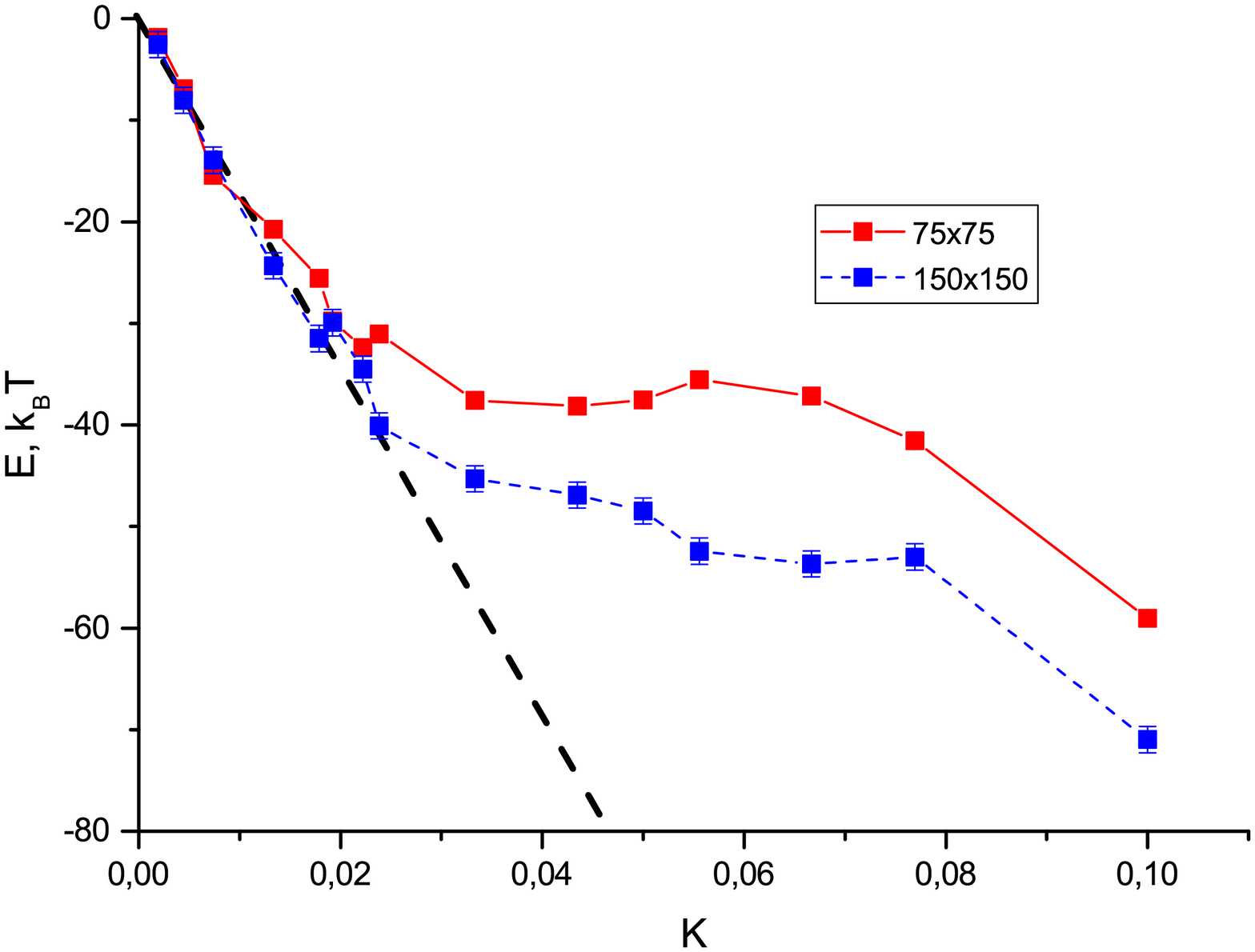}
		\caption{The energy of a defect curve of fixed length $L = 75$ for two lattice sizes: $75\times75$ and $150\times150$. The volume effect of repulsion is weakened. Data points are connected by line for clarity.}
		\label{fig:curved_difflatenergy}
	\end{figure}

	\figurename~\ref{fig:curved_energy1} shows the dependence of the specific energy on the curvature $K = 1/R$. Considering the region of low curvatures, we note that the specific energy is almost the same for all lattice sizes. The reason is that the curve is almost like a straight line and extends from the edge to the edge of the grid. By dividing the total energy by the number of nodes, we can not distinguish two lines along their length (since we deal with the energy per unit length). For larger curvatures, the difference is significant, because the number of spin nodes between ``semicircular branches'' depends on the size of the lattice. At the same time, the graphs in \figurename~\ref{fig:curved_energy1} has a feature in behavior: there is a region of local maximum where the energy grows with the curvature. This does not correspond to the prediction (\ref{eq:avecircleintenergy}), which asserts a monotonous decrease. Since this formula is derived from physical considerations, it can be argued that there is no physical cause for energy growth, i.e. repulsion between branches.
	
	An attentive reader will note that for line of length $L=100$ the maximal possible curvature (when line is folded into a circle) is K=0.063, but \figurename~\ref{fig:curved_energy1} has a point $K=0,066...$ corresponding $R=15$. This illustrates the problem of the discretization algorithm and the defect line put in lattice is not a perfect arc of a circle at all.
	
\begin{figure}[h]
		\centering
		\includegraphics[width=0.6\linewidth]{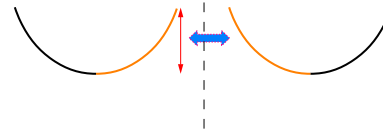}
		\caption{Attraction between two curved lines arises then two adjacent endings become significant in perpendicular dimension. Red arrow shows a part of the arc visible from the side of the other arc copy. A dashed line indicates a border with periodic boundary conditions.}
		\label{fig:curvesattraction}
	\end{figure}

	Let's analyze the curvature value $K = 0.035$ (where the curve rise in \figurename~\ref{fig:curved_difflatenergy} starts) for the lattice $100\times100$. The radius corresponding to this curvature is
	\begin{equation}
	\label{eq:radius}
	R=\frac1K\approx 28.571.
	\end{equation}
	A circle whose half-length is $L_{1/2} = 100$ has a diameter
	$$
	d = \frac{2L_{1/2}}{\pi}\approx 63.662,
	$$
	which is approximately twice the size of $R$ (see (\ref{eq:radius})). This means that the curve with this curvature has the form of a semicircle. Due to periodic boundary conditions, many identical copies of the defect curve are located in the system at a certain distance from each other. And they are attracted to each other. With further bending, the branches of each circle approach each other, but move away from the corresponding branches of adjacent curves. This increases the energy of the system. Consequently, the observed local maximum is a \textbf{volume effect}. \figurename~\ref{fig:curvesattraction} demonstrates the phenomenon: in case of small $K$ only the endings themselves take part in the interaction, so few spins are involved in the process. But after the line is distorted, a lot more nodes appear opposite their counterparts, so energy volume effect is tangible. Finally, as the curvature advances, the intensity of the effect decreases, and the attraction again lowers the energy.
	
	To overcome this problem, it would be good to move the neighboring curves far away from each other (to an infinite distance), but this is not possible in a lattice model. In this case, at least we try to double the size of the lattice: replace the lattice of $75\times75$ by $150\times150$. The length of the curve remains the same: $L =  75$. As before, the relative energy was calculated. The data obtained are shown in \figurename~\ref{fig:curved_difflatenergy}. The solid line has a local maximum (the region of repulsion of the branches), but the dashed line is almost monotonic and more adjacent to the linear approximation of the region of small curvature. In the second case, the neighboring curves interact less (due to the greater distance between them) and the attraction between the branches of one curve prevails.
	
	\begin{figure}[h]
		\centering
		\includegraphics[width=0.8\linewidth]{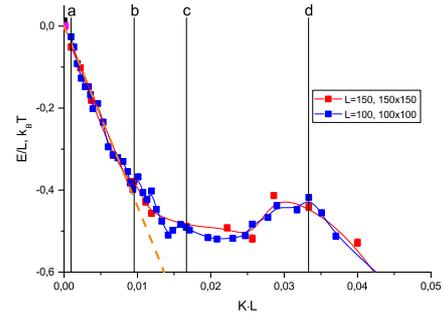}
		\caption{Manifestation of scale invariance of the specific energy $E/L$ of the defective curve at $1/T=0.44$. Both axes are dimensionless. B-Splines are used for clarity. The curve's form in four sections (a - d) are shown in \figurename~\ref{fig:curvecollapsing}.}
		\label{fig:curved_energyinv}
	\end{figure}
	
	\begin{figure}[h]
		\centering
		\includegraphics[width=1\linewidth]{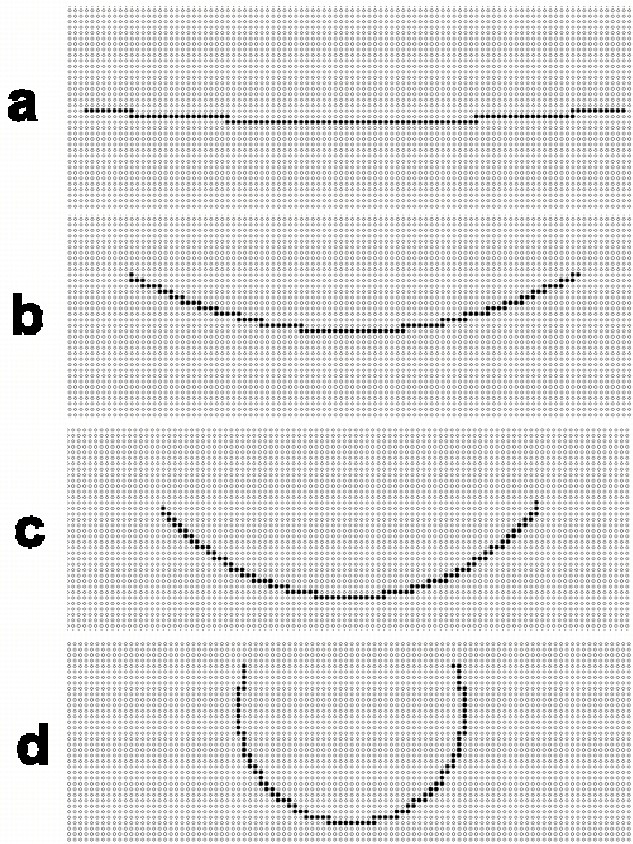}
		\caption{ The numerical snapshots of a curve $L=100$ collapsing with increasing $K$. The relevant values of $K\cdot L$ are shown in \figurename~\ref{fig:curved_energyinv}. Black nodes are defects while the others are spins.}
		\label{fig:curvecollapsing}
	\end{figure}
	
	To study the scale invariance, we draw attention to the fact that the curvature used in \figurename~\ref{fig:curved_energy1} has the dimension of inversed length. Let us plot the dimensionless product $KL$ along the horizontal axis. The result is shown in \figurename~\ref{fig:curved_energyinv}. The graphs coincide with remarkable accuracy. This agrees with the existence of scale invariance in the vicinity of the phase transition. It is important to note that the local maximum of the observed curves is a manifestation of the volume effect. The study showed that when considering large lattice, the linear energy's dependence on curvature is strongly pronounced. According to the graphs presented in \figurename~\ref{fig:curved_energyinv}, the coefficient of this dependence is defined:
	\begin{equation}
	\label{eq:specenergyscinv}
	\frac EL=aKL,\quad a = -43,9 \pm 0,9.
	\end{equation}
	
	Let us calculate the specific energy of a flat helix analog (that is a circle considered in our research). The curve discussed above should be bound into a circle, so that $L/R = 2\pi$. Using (\ref{eq:specenergyscinv}), we get the Casimir's energy of helix, referred to its length $E/L=2\pi a$. In this case, the energy required to turn a helix into a line equals $2\pi aL$. We suggest this to be the physical interpretation of the constant $a$.

	\section{Summary and conclusion}
	
	The aim of our work was to study the collapse of defect structures in the 2D Ising spin model. We focus our intention on the problem of Casimir interaction of defects. To study this problem we used the numerical Monte-Carlo simulation of Ising model with defects.
		
	An evolution of 4 defect-length line was studied from the viewpoint of the most probable direction of state change. We found out that the internal energy of 4 defects line has its maximum at critical point (temperature of phase transition). The simplest method of deformation is proposed, and considerations were given about the energy release. Using the direct Monte-Carlo simulation, the energy path of the reaction was found. The presence of a metastable state predicted by analyzing the reaction stages was confirmed. The existence of a number of energetic thresholds was established. The effect was discussed at three different temperatures. The comparison revealed that the collapse is the most beneficial at the phase transition point.
	
	
	A study of the Casimir forces of interaction between extensive objects in spin models can help in studying the folding process of proteins. As noted in the Introduction, water can be described as a statistical system within a framework of a lattice gas model, and a protein --- as a configuration of vacancies. In the 2D Ising model a protein is described as a defect line. As a simplification, the entire variety of possible curves of fixed length was limited to curves of constant curvature. The one-parameter family of curves obtained in this way was investigated from the point of view of the most energetically favorable state. A tendency to radius decreasing is observed (is looks like a collapse to a circle). Monte Carlo simulation showed that the reduction of the Casimir energy of such a line is proportional to its curvature. This indicates that the critical Casimir effect leads to an increase of curvature and thus facilitates the folding. The dependence of the effect on the size of the lattice containing the curve is checked. The origin of the local maximum of the energy's dependence on the curvature is explained. A graph of normalized quantities  in order to eliminate the volume effect is depicted.
	
	\section{Acknowledgements}
	
	The reported study was supported by the Supercomputing Center of Lomonosov Moscow State University \cite{parallelru}. This work has been supported by the grant from the Russian Science Foundation (project number 16-12-10059). The authors gratefully thank to the Referees for the constructive comments and recommendations.
	
	\bibliography{curveddef}

\begin{thebibliography}{10}

\bibitem{nanoribbons}
Leonardo~C. Campos, Vitor~R. Manfrinato, Javier~D. Sanchez-Yamagishi, Jing
  Kong, and Pablo Jarillo-Herrero.
\newblock Anisotropic etching and nanoribbon formation in single-layer
  graphene.
\newblock {\em Nano Letters}, 9(7):2600--2604, 2009.
\newblock PMID: 19527022.

\bibitem{fisher}
M.~E. Fisher and P.-G. de~Gennes.
\newblock Wall phenomena in a critical binary mixture.
\newblock {\em Comptes rendus de l'Académie des sciences (Paris) Ser. B},
  287:207--209, 1978.

\bibitem{krech}
M.~Krech.
\newblock {\em The {C}asimir Effect in Critical Systems}.
\newblock 1994.

\bibitem{gamb}
Andrea Gambassi.
\newblock The {C}asimir effect: From quantum to critical fluctuations.
\newblock {\em Journal of Physics: Conference Series}, 161(1):012037, 2009.

\bibitem{kadanoff}
L.~P. Kadanoff.
\newblock {Scaling laws for {I}sing models near T(c)}.
\newblock {\em Physics}, 2:263--272, 1966.

\bibitem{selke1}
Walter Selke.
\newblock The two-dimensional {I}sing model with two defect lines a monte carlo
  study.
\newblock {\em Physica A: Statistical Mechanics and its Applications},
  177(1):460 -- 465, 1991.

\bibitem{selke2}
W.~Selke, N.~M. {\v{S}}vraki{\'{c}}, and P.~J. Upton.
\newblock {I}sing models with interfaces, defect lines, and walls.
\newblock {\em Zeitschrift f{\"u}r Physik B Condensed Matter}, 89(2):231--237,
  Jun 1992.

\bibitem{hasen}
Martin Hasenbusch.
\newblock Thermodynamic {C}asimir forces between a sphere and a plate: Monte
  carlo simulation of a spin model.
\newblock {\em Phys. Rev. E}, 87:022130, Feb 2013.

\bibitem{nowakowski}
P~Nowakowski, A~Maciolek, and S~Dietrich.
\newblock Critical {C}asimir forces between defects in the 2d {I}sing model.
\newblock {\em Journal of Physics A: Mathematical and Theoretical},
  49(48):485001, 2016.

\bibitem{hecht}
Robert Hecht.
\newblock Correlation functions for the two-dimensional {I}sing model.
\newblock {\em Phys. Rev.}, 158:557--561, Jun 1967.

\bibitem{cells}
Benjamin~B. Machta, Sarah~L. Veatch, and James~P. Sethna.
\newblock Critical {C}asimir forces in cellular membranes.
\newblock {\em Phys. Rev. Lett.}, 109:138101, Sep 2012.

\bibitem{EB}
Theodore~W. Burkhardt and Erich Eisenriegler.
\newblock Casimir interaction of spheres in a fluid at the critical point.
\newblock {\em Phys. Rev. Lett.}, 74:3189--3192, Apr 1995.

\bibitem{bell1}
G~M Bell.
\newblock Statistical mechanics of water: lattice model with directed bonding.
\newblock {\em Journal of Physics C: Solid State Physics}, 5(9):889, 1972.

\bibitem{bell2}
G.~M. Bell and D.~W. Salt.
\newblock Three-dimensional lattice model for the water/ice system.
\newblock {\em J. Chem. Soc.{,} Faraday Trans. 2}, 72:76--86, 1976.

\bibitem{titov}
S.~V. Titov and Yu.~K. Tovbin.
\newblock A molecular model of water based on the lattice gas model.
\newblock {\em Russian Journal of Physical Chemistry A}, 85(2):194--201, Feb
  2011.

\bibitem{nanopores}
Jurgen Kofinger, Gerhard Hummer, and Christoph Dellago.
\newblock Single-file water in nanopores.
\newblock {\em Phys. Chem. Chem. Phys.}, 13:15403--15417, 2011.

\bibitem{lee}
T.~D. Lee and C.~N. Yang.
\newblock Statistical theory of equations of state and phase transitions. ii.
  lattice gas and {I}sing model.
\newblock {\em Phys. Rev.}, 87:410--419, Aug 1952.

\bibitem{os}
Takeo Osawa and Katuro Sawada.
\newblock Critical properties of {I}sing models containing dilute impurities.
\newblock {\em Progress of Theoretical Physics}, 49(1):83--88, 1973.

\bibitem{ha}
A~B Harris.
\newblock Effect of random defects on the critical behaviour of {I}sing models.
\newblock {\em Journal of Physics C: Solid State Physics}, 7(9):1671, 1974.

\bibitem{china}
Xintian Wu and Yangyang Zhang.
\newblock Critical phenomena of a single defect.
\newblock {\em Phys. Rev. E}, 92:032108, Sep 2015.

\bibitem{chboard}
Tobias Preis, Peter Virnau, Wolfgang Paul, and Johannes~J. Schneider.
\newblock {GPU} acceelerated {M}onte {C}arlo simulation of the 2{D} and 3{D}
  {I}sing model.
\newblock {\em Journal of Computational Physics}, 228(12):4468--4477, 2009.

\bibitem{critCasTemp}
Lars Onsager.
\newblock Crystal statistics. i. a two-dimensional model with an order-disorder
  transition.
\newblock {\em Phys. Rev.}, 65:117--149, Feb 1944.

\bibitem{parallelru}
Sobolev S.I. Antonov A.S. Bryzgalov P.A. Nikitenko D.A. Stefanov K.S.
  Voevodin~Vad.V Voevodin~Vl.V., Zhumatiy~S.A.
\newblock Practice of "{L}omonosov" {S}upercomputer.
\newblock {\em Open Syst. J.}, pages 36--39, 2012.

\end{thebibliography}
	\bibliographystyle{unsrt}

\end{document}